\documentclass[twocolumn,aps,prl]{revtex4}

\usepackage{amsmath}
\usepackage{amsfonts}
\usepackage{amssymb}
\usepackage[colorlinks=true,citecolor=blue,linkcolor=red]{hyperref}
\usepackage{graphicx}
\usepackage{bbold}					
\usepackage[dvipsnames]{xcolor}

\usepackage{ulem}
\usepackage{bm} 
\usepackage{subfigure}
\usepackage{dcolumn}
\usepackage{mathtools}

\usepackage{array,booktabs,ragged2e}	
\usepackage{soul}						
\usepackage{cancel}						
\usepackage{tabularx}					
\usepackage{multirow}
\usepackage{hhline}
\usepackage{adjustbox}
\usepackage{sidecap}

\usepackage[utf8]{inputenc}
\usepackage[T1]{fontenc}

\newcommand{\angstrom}{\mbox{\normalfont\AA}}	

\def\bf#1{\textbf{#1}}

\newcommand{\bk}{\pmb{k}}

\newcommand{\bsigma}{\pmb{\sigma}}

\newcolumntype{P}[1]{>{\raggedleft\arraybackslash}p{#1}}
\newcolumntype{R}[1]{>{\centering\arraybackslash}p{#1}}

\begin{document}


\title{Chiral Decomposition of Twisted Graphene Multilayers with Arbitrary Stacking}

\author{ShengNan Zhang$^{1,2}$}
\altaffiliation{These authors contributed equally}
\author{Bo Xie$^{3}$}
\altaffiliation{These authors contributed equally}
\author{QuanSheng Wu$^{1,2}$}
\email{quansheng.wu@epfl.ch}
\author{Jianpeng Liu$^{3,4}$} 
\author{Oleg V. Yazyev$^{1,2}$}
\email{oleg.yazyev@epfl.ch}

\affiliation{$^{1}$Institute of Physics, Ecole Polytechnique F\'{e}d\'{e}rale de Lausanne (EPFL), CH-1015 Lausanne, Switzerland}
\affiliation{${^{2}}$National Centre for Computational Design and Discovery of Novel Materials MARVEL, Ecole Polytechnique F\'{e}d\'{e}rale de Lausanne (EPFL), CH-1015 Lausanne, Switzerland}
\affiliation{${^{3}}$School of Physical Sciences and Technology, ShanghaiTech University, Shanghai 200031, China}
\affiliation{${^{4}}$ShanghaiTech laboratory for topological physics, ShanghaiTech University, Shanghai 200031, China}
\date{\today}

\begin{abstract}
We formulate the chiral decomposition rules that govern the electronic structure of a broad family of twisted $N+M$ multilayer graphene configurations that combine arbitrary stacking order and a mutual twist.
We show that at the magic angle in the chiral limit the low-energy bands of such systems are composed
of chiral pseudospin doublets which are energetically entangled with two flat bands per valley induced by the moir\'e superlattice potential.
The analytic analysis is supported by explicit numerical calculations based on realistic parameterization. We further show that applying vertical displacement fields can open up energy gaps between the pseudospin doublets and the two flat bands, such that the flat bands may carry nonzero valley Chern numbers.
These results provide guidelines for the rational design of various topological and correlated states in generic twisted graphene multilayers. 
\end{abstract} 

\maketitle
     

Graphene moir\'e superlattices~\cite{Lopes2007,Bistritzer2011} have recently emerged as a novel platform for exploring correlated physics~\cite{Cao2018a,Chen2019,Lu2019,Burg2019,Shen2020}, energy band topology~\cite{Zou2018,Song2019,Po2019,Ahn2019,Serlin2020} 
and superconductivity~\cite{Cao2018b,Yankowitz2019,Lu2019}. A moir\'e superlattice is formed upon combining two layers of two-dimensional (2D) materials with lattice mismatch or relative twist of their lattices. On one hand, the moir\'e superlattice potential alters the electronic structure, e.g. flattening the energy bands to only a few meV bandwidth at a certain value of twist angle~\cite{Bistritzer2011}, that is responsible for the recently observed superconducting and correlated insulator states \cite{Cao2018a, Cao2018b}. 
On the other hand, the moir\'e superlattices generate pseudo-gauge fields leading to non-trivial band topology~\cite{San-Jose2012, Liu2019a, Liu2019,Tarnopolsky2019}, 
which in turn results in the Chern insulator phase~\cite{Wu2020,Das2020, Polshyn2020, Saito2020, Shen2020b} and the quantum anomalous Hall effect~\cite{Serlin2020}. The interplay between superconductivity, electron correlation and band topology make the moir\'e superlattices a marvel of condensed matter physics. 

Following the breakthrough works on twisted bilayer graphene (TBG) \cite{Cao2018a,Cao2018b}, the scope of investigations has quickly extended to other graphene moir\'e superlattices including twisted double bilayer graphene (TDBG)~\cite{Burg2019,Haddadi2019,Chebrolu2019,Koshino2019,Shen2020,Rickhaus2019}, ABC graphene/hexagonal boron nitride (hBN)~\cite{Chittari2019,Chen2019,Chen2019b}, twisted mono-bilayer graphene~\cite{Rademaker2020,Polshyn2020}, twisted $N+M$ graphene multilayers based on ABC-~\cite{Liu2019} and AB-stacking~\cite{Ma2020}, more complex arrangements~\cite{Khalaf2019,Zhu2020,Lopez-Bezanilla2020,Lei2020,Carr2020,Park2020,Hao2020}, and even infinite stacks of twisted layers~\cite{Cea2019}. 

\begin{figure}[b]
\begin{center}
    \includegraphics[width=8cm]{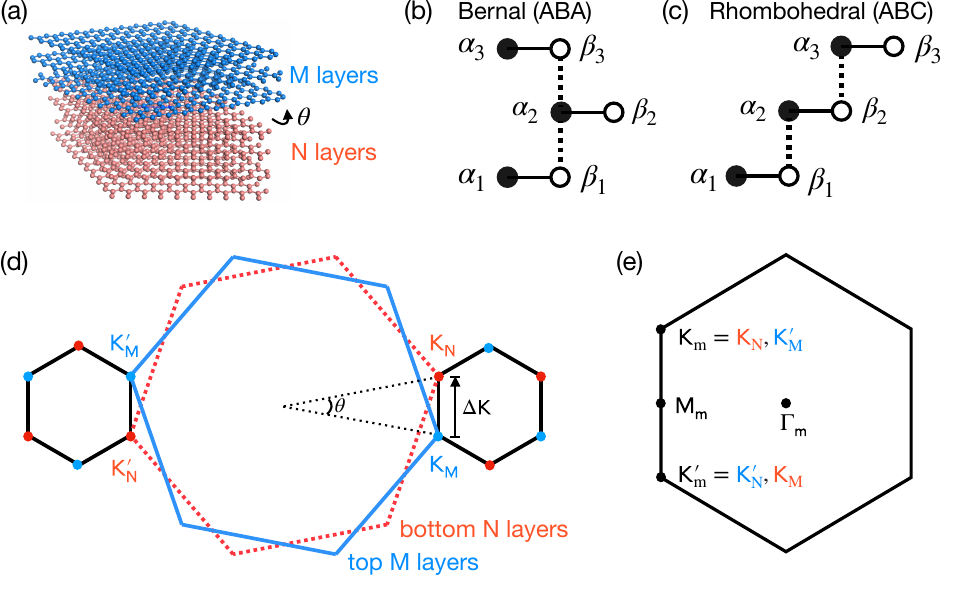}
\caption{
(a) Schematic representation of twisted $N+M$ multilayer graphene, and (b) Bernal and (c) rhombohedral stacking configurations. 
(d) The Brillouin zones of top (blue) and bottom (red) ordered multilayer graphene counterparts. 
Black hexagons represent the moir\'e reciprocal lattices at valleys $K$ and $K'$. 
(e) Moir\'e Brillouin zone plotted with relevant high-symmetry points. }
\label{fig_BZ}
\end{center}
\end{figure}

In this work, we study a broadly defined class of twisted $N+M$ multilayer graphene configurations, in which an ordered $N$-layer graphene is twisted with respect to another ordered $M$-layer graphene by angle $\theta$ (Fig.~\ref{fig_BZ}(a)). The two ordered counterparts, below referred to as top and bottom, can  assume energetically favorable Bernal (ABA), rhombohedral (ABC) or mixed stacking sequences, and thus called mixed multilayer graphene (MMG) hereafter. 
Figs.~\ref{fig_BZ}(b) and (c) illustrate the Bernal and rhombohedral sequences with $\alpha$ and $\beta$ sublattices within each monolayer being distinguished.
Considering only the nearest-neighbour hoppings, the low-energy electronic structure of MMGs in the chiral limit consists of chiral pseudospin doublets with a conserved chirality sum \cite{Min2008a,Min2008b,Koshino2013}. The bands of each doublet are described by a pseudospin (denoting sublattice) Hamiltonian 
\begin{equation}
    \mathcal{H}_J(\bk)\propto k^J [{\rm{cos}}(J\phi_{\bk}){\sigma}_x\pm {\rm{sin}}(J\phi_{\bk}){\sigma}_y] ,
    \label{eqn:doublet}
\end{equation}
where $\bk$ is the momentum measured from valleys $\pmb{K}$ or $\pmb{K'}$, ${\sigma}_x$, ${\sigma}_y$ are the Pauli matrices acting on graphene sublattices, $k=|\bk|$, $\phi_{\bk}$ denotes the orientation of $\bk$, and $\pm J$ is the chirality index of a doublet in graphene valleys $K$ and $K'$. The low-energy effective Hamiltonian of an $N$-layer MMG can be decomposed as follows
\begin{equation}
    \mathcal{H}_N^{eff}\sim H_{J_1}\oplus H_{J_2}\oplus \cdots \oplus H_{J_{N_D}}\label{Eq:HN_decompose} ,
\end{equation}
where $N_D$ is the number of doublets that depends on the details of the stacking sequence. The chirality indices $\{J_i\}$ fulfill 
\begin{equation}
    \sum_{i=1}^{N_D}J_i=N .
\end{equation}

The low-energy band of an MMG system described by Eq.(\ref{Eq:HN_decompose}) can be read off by partitioning a stack according to the following rules reproduced from Refs.~\onlinecite{Min2008a,Min2008b} for completeness: (i) Identify the longest non-overlapping segments within which there are no reversals of the stacking sense. When there is an ambiguity in the selection of non-overlapping segments, choose the partitioning which incorporates the largest number of layers. (ii)  Partition iteratively the remaining segments of the stack into smaller $J$ elements, excluding layers contained within previously identified partitions, until all layers are exhausted. 
Each segment 
defines a $J$-layer partition of the stack and is associated with a chirality $J$ doublet.
The low-energy bands of $N+M$-layer MMG can be interpreted using these rules. In our work, we derive partitioning rules that extend this approach to twisted $N+M$-layer MMG (TMMG). 

Before proceeding, we define the notations in our TMMG partitioning scheme. The bottom $N$-layer (top $M$-layer) multilayers are divided into $N_D$ ($M_D$) segments, the lengths of which are $J_1, J_2, \cdots, J_{N_D}$ ($J'_1, J'_2, \cdots, J'_{M_D}$) from bottom to the top. Segments $J_{N_D}$ and $J'_1$ which are mutually twisted are referred to as the moir\'e segments. The generic partitioning rules for TMMG are hereby given as: (i) $N+M$ layers are divided into $N_D+M_D-2$ common segments $J_1, J_2, \cdots, J_{N_D-1}, J'_2, J'_3, \cdots, J'_{M_D}$ and two moir\'e segments $J_{N_D}$ and $J'_1$. (ii) Choose the longest segment for $J_{N_D}$ ($J'_1$) if there is ambiguity between $J_{N_D}$ and $J_{N_D-1}$ ($J'_{1}$ and $J'_{2}$) (see the examples in Table~\ref{tab:partitioning}). Eventually, the low-energy bands comprise two flat bands per valley originating from the twisted $J_{N_D}$+$J'_{1}$ 
rhombohedral stacking segments, $N_D-1$ doublets at $K_m$ ($K_m'$) of $K$ ($K'$) valley and $M_D-1$ doublets at $K_m'$ ($K_m$) of $K$ ($K'$) valley from the other segments.  The chiralities of these doublets are $J_1$, $J_2$,...,$J_{N_D-1}$,  $J'_1$, $J'_2$,...,$J'_{M_D-1}$ if $J_{N_D}>1$ and $J'_1>1$. 
Exceptionally,
the chirality of the doublet originating from $J_{N_D-1}$ ($J'_2$) segment, next to the moir\'e layers, is $J_{N_D-1}-1$ ($J'_2-1$) if $J_{N_D}=1$ ($J'_{1}=1$).

To apply these rules, we begin with the simplest cases of twisted rhombohedral  stacking multilayer graphene (TRMG), e.g. A-AB, AB-AB, A-ABC, AB-ABC and ABC-ABC, since no stacking sequence reversal takes place in the top and bottom counterparts of these configurations. According to rule (i), there are only two moir\'e segments, i.e. $N_D=M_D=1$. Band structures of these configurations are very similar to each other with two flat bands per valley isolated from the remote bands by energy gaps as in the TBG~\cite{Liu2019}.

The most complex case is the twisted Bernal stacking multilayer graphene (TBMG) owing to stacking sequence reversal as in the examples given in Table~\ref{tab:partitioning}. 
For $N,M<3$, TBMG systems are equivalent to TRMGs. For $N\le2$ and $M\ge3$ ($N\ge3$ and $M\le2$), the length of moir\'e segments is fixed to  $J_1=N$, $J'_1=2$ ($J_1=2$,  $J_1'=M$). 
The low-energy spectrum is composed of 2 flat bands per valley resulting from a twisted $N+2$ ($M+2$) graphene and $(M-2)/2$ ($(N-2)/2$) doublets of chirality $J=2$  with even $M$ ($N$); when $M$ ($N$) is odd, an extra doublet of chirality $J=1$ in addition to the $(M-3)/2$ ($(N-3)/2$) $J=2$ doublets at $K_m'$ ($K_m$) of valley K and at $K_m$ ($K_m'$) of valley K'. 
For $N,M \ge 3$, $J_{N_D}=2$, $J'_1=2$, and the low-energy spectrum is composed with 2 flat bands per valley resulting from the TDBG moir\'e segments at the twist interface and $(M-2)/2 ((N-2)/2)$ doublets of chirality $J=2$ for even $N$ and $M$ at $K_m'$ ($K_m$) of valley K and at $K_m$ ($K_m'$) of valley K'. In the case of odd $N$ ($M$) the bottom and top counterparts contribute $(N-3)/2$ ($(M-3)/2$)
$J=2$ doublets and one $J=1$ doublet. 
Our conclusions are in agreement with the numerical results of Ref.~\onlinecite{Ma2020}.

Between the TBMG and TRMG limits, intermediate TMMG configurations show a rich variety of low-energy behaviors which can be predicted using the above-mentioned rules. Taking ABA-ABCBC \footnote{The layers left to right correspond to bottom to top, '-' indicates the twist.} as an example, BA-ABC is identified as a $2+3$ TRMG, the remaining A on the bottom layer gives a $J=1$ doublet at $K_m$ ($K_m'$) of valley $K$ ($K'$) and the remaining BC layers gives a $J=2$ doublet at $K_m'$ ($K_m$) of valley $K$ ($K'$). The low-energy band structure of this TMMG configuration has two sets of low-energy bands with $|E|\propto k$, $k^2$ and four flat bands.

\begin{table}[]
    \centering
    \caption{Examples of chirality decomposition of TMMG configurations. Two flat bands per valley generated by the moir\'e segments (in bold) are not counted, while $\chi=0$ indicates there are no other energy level besides the flat bands. Band structures of these systems are given in the Supplementary Material.  }
    \begin{tabular}{l|l|c|c}
    \hline\hline
      configuration   & partitioning & $\chi_K (\rm K_m)$ & $\chi_K (\rm K_m')$ \\
          &   & $\chi_{K'} (\rm K_m')$ & $\chi_{K'} (\rm K_m)$ \\
      \hline
       A-ABA  & \bf{A-AB}+A & 0 & 1 \\
       A-ABAB  & \bf{A-AB}+AB & 0 & 2 \\
       AB-ABA  & \bf{AB-AB}+A & 0 & 1 \\
       ABA-ABA  & A+\bf{BA-AB}+A & 1 & 1 \\
       A-ABAC  & \bf{A-A}+BAC & 0 & 2 \\
       ABA-ABCBA &A+\bf{BA-ABC}+BA & 1 & 2 \\
       ABA-ABCBC  &A+\bf{BA-ABC}+BC  & 1 & 2\\
       ABA-ABCBAB  &A+\bf{BA-ABC}+BA+B  & 1 & 2+1\\
       \hline\hline
    \end{tabular}
    \label{tab:partitioning}
\end{table}

In order to understand and verify the partitioning rules, we adopt the continuum model introduced by Bistritzer and MacDonald for deriving the effective Hamiltonian of TMMG. As introduced in Ref.~\onlinecite{Liu2019}, the flat bands of a TBG system originate from the zeroth pseudo Landau levels (LL)~\cite{Liu2019} under certain gauge transformations. In particular, the energies of the zeroth pseudo LLs are exactly zero in the chiral limit at the magical angle~\cite{Tarnopolsky2019}, i.e. when the Fermi velocity of the flat bands is exactly zero. Here, the chiral limit implies that the coupling in the AA region of the moir\'e superlattice is set to zero and only the nearest-neighbour interlayer coupling is considered. We will discuss the low-energy physics at the magic angle in the chiral limit and at valley $K$ hereafter, unless other conditions are explicitly specified. Following the strategy of Ref.~\onlinecite{Liu2019}, we can divide the system into three parts: $(N-1)$ layers MMG, $(M-1)$ layers MMG and a TBG subsystem. 
Using the wave functions of two zeroth pseudo LLs |LL$_{0,\alpha}\rangle$,  |LL$_{0,\beta}\rangle$ from TBG layers and the basis of the other two parts, the low-energy effective Hamiltonian of the TMMG~\footnote{In the main text, we only show the Hamiltonian for the TMMG in which the chirality of the top two layers of bottom part and the bottom two layers of top part are the same. Other situations are discussed in the Supplementary Material.} of the $K$ valley and $K_m$ is expressed as~\cite{supp}
\begin{equation}
\tilde{H}_K^{K_m}(N+M)=\begin{pmatrix}
h_b(\bk) & 0 \\
0 & h_t(\bk) 
\end{pmatrix}\;,
\label{eq:HMN}
\end{equation}
where $h_b(\bk)$ and $h_t(\bk)$ are the effective Hamiltonians for the $N$-layer and $M$-layer MMGs. In particular, 
\begin{equation}
h_b(\bk)=\begin{pmatrix}
 & \cdots &  &   \\
 \cdots & h_0(\bk) &  h_{\nu_{N-2}} &0 \\
  & h^{\dagger}_{\nu_{N-2}} & h_0(\bk) &  T_L^{b\dagger}\\
 & 0 & T_L^b & 0 &   
\end{pmatrix} \;
\label{eq:HM}
\end{equation}
and
\begin{equation}
h_t(\bk)=\begin{pmatrix}
0 & T_L^t & 0 &   \\
T_L^{t\dagger} & h_0^{\Delta}(\bk) &  h_{\nu_{N+2}} &  \\
    0 &  h^{\dagger}_{\nu_{N+2}} & h_0^{\Delta}(\bk) &  \cdots\\
 &  &\cdots  &  
\end{pmatrix} \;
\label{eq:HM},
\end{equation}
where  $h_{0}(\bk)\!=\!-\hbar v_{F}\bk\cdot\bsigma$ and  $h_{0}^{\Delta}(\bk)\!=\!-\hbar v_{F}(\bk+\Delta K)\cdot\bsigma$ stand for the low-energy effective Hamiltonians for monolayer graphene of the bottom and top layers near the Dirac point $\mathbf{K}_N$, $\bk$ is measured from $\mathbf{K}_N$, $\Delta K=\pmb{K_N}-\pmb{K_M}$, $h_{\nu_i}$ is the interlayer hopping, with
\begin{equation}
h_{+}=\begin{pmatrix}
0 & 0\;\\
t_{\perp} & 0 
\end{pmatrix}\;,
\label{eq:chiral-hopping}
\end{equation}
and $h_{-}=h_{+}^{\dagger}$. $\nu_i=+1$ if the stacking sequence is AB, BC, CA, $\nu_i=-1$ otherwise. $T^{t,b}_L$ is the coupling matrix between a zeroth pseudo Landau level and the graphene layer adjacent the TBG layers. 
$T^t_L=(t_l, 0)$ and $T^b_L=(0,t_l)$ when the stacking chirality is 1, $T^t_L=(0, t_l)$ and $T^b_L=(t_l, 0)$ if it is $-1$. $t_l$ is the coupling coefficient.

Without loss of generality, we choose two typical configurations A-ABA and A-ABAC and derive the low-energy effective Hamiltonians in order to demonstrate the partitioning rules. 
For the A-ABA system, enumerating the layers from left to right, we devise the following basis set \{|LL$_{0,\alpha}\rangle$,  |LL$_{0,\beta}\rangle$, $|\alpha_2\rangle$, $|\beta_2\rangle$, $|\alpha_3\rangle$, $|\beta_3\rangle$, $|\alpha_4\rangle$, $|\beta_4\rangle$, $|\alpha_5\rangle$, $|\beta_5\rangle$\}, in which the effective Hamiltonian of $K$ valley at $K_m$  
\begin{equation}
H_+(\bk)=\begin{pmatrix}
0 & 0 & 0 & 0 \\
0 & 0 & T_L^t & 0  \\
0 & T_L^{t\dagger} & h_0^{\Delta}(\bk) &  h_-  \\
0 & 0 &  h_+ & h_0^{\Delta}(\bk)
\end{pmatrix}\label{eq:HABA},
\end{equation}
where $T^t_L=(t_l, 0)$. The effective Hamiltonian of $K$ valley at $K_m'$ is 
\begin{equation}
H_-(\bk)=\begin{pmatrix}
0 & 0 & 0 & 0 \\
0 & 0 & T_L^t & 0  \\
0 & T_L^{t\dagger} & h_0(\bk) &  h_-  \\
0 & 0 &  h_+ & h_0(\bk)
\end{pmatrix} \;
\label{eq:HABAprime}.
\end{equation}
The low-energy spectrum of $H_+(\bk)$ 
consists of two dispersionless zero-energy bands and four remote high-energy bands. The local density of states (LDOS) of these two zero energy levels is distributed over the $\alpha$ sublattice of TBG layers A-A and the $\beta$ sublattice of the second layer B of the ABA counterpart.

To obtain the low-energy effective Hamiltonian of Eq.(\ref{eq:HABAprime}), we apply degenerate-states perturbation theory to the zero-energy subspace \cite{Sakurai2017,Min2008a}, which includes $|LL_{0,\alpha}\rangle$, two isolated site states $|\beta_3\rangle$ and $|\alpha_4\rangle$, and a zero energy state $|\beta^-_4\rangle$ resulting from the hybridization of $|LL_{0,\beta}\rangle$, $|\alpha_3\rangle$, $|\beta_4\rangle$ as
\begin{eqnarray}
|\beta^-_4\rangle=\frac{t}{\sqrt{t^2+|t_l|^2}} (|LL_{0,\beta}\rangle-\frac{|t_l|}{t}|\beta_4\rangle) . \label{eqn:beta4minus}\end{eqnarray}
First-order perturbation theory gives
\begin{eqnarray}
\langle\alpha_4|H|\beta^-_4\rangle=\frac{t_l}{\sqrt{t^2+|t_l|^2}}\hbar v_F k_- ,
\end{eqnarray}   
and accordingly we obtain the low-energy effective Hamiltonian of $H_-(\bk)$ in the basis of \{\{$|LL_{0,\alpha}\rangle$,  $|\beta_3\rangle$\}, \{$|\alpha_3\rangle$, $|\beta_4^-\rangle$\}\} as
\begin{equation}
H_-^{eff}(\bk)=\begin{pmatrix}
0 & 0 \\
0 & 0\end{pmatrix} \oplus
\begin{pmatrix}
0 & \hbar \tilde{v}_F k_- \\
\hbar \tilde{v}_F k_+ & 0
\end{pmatrix} \;
\label{eq:HABAprime_eff} .
\end{equation}
where $k_{\pm}=k_x\pm+i k_y$. The zero matrix in Eq.(\ref{eq:HABAprime_eff}) gives two flat bands, and their LDOS is distributed only on the $\alpha$ sublattice of TBG layers A-A and the $\beta$ sublattice of the second layer B of the top stacking ABA. The second matrix in Eq.(\ref{eq:HABAprime_eff}) describes the well-known $J=1$ massless Dirac equation with a reduced Fermi velocity $\tilde{v}_F\sim\frac{|t_l|}{\sqrt{t^2+|t_l|^2}}v_F$.

\begin{figure}
\begin{center}
\includegraphics[width=8cm]{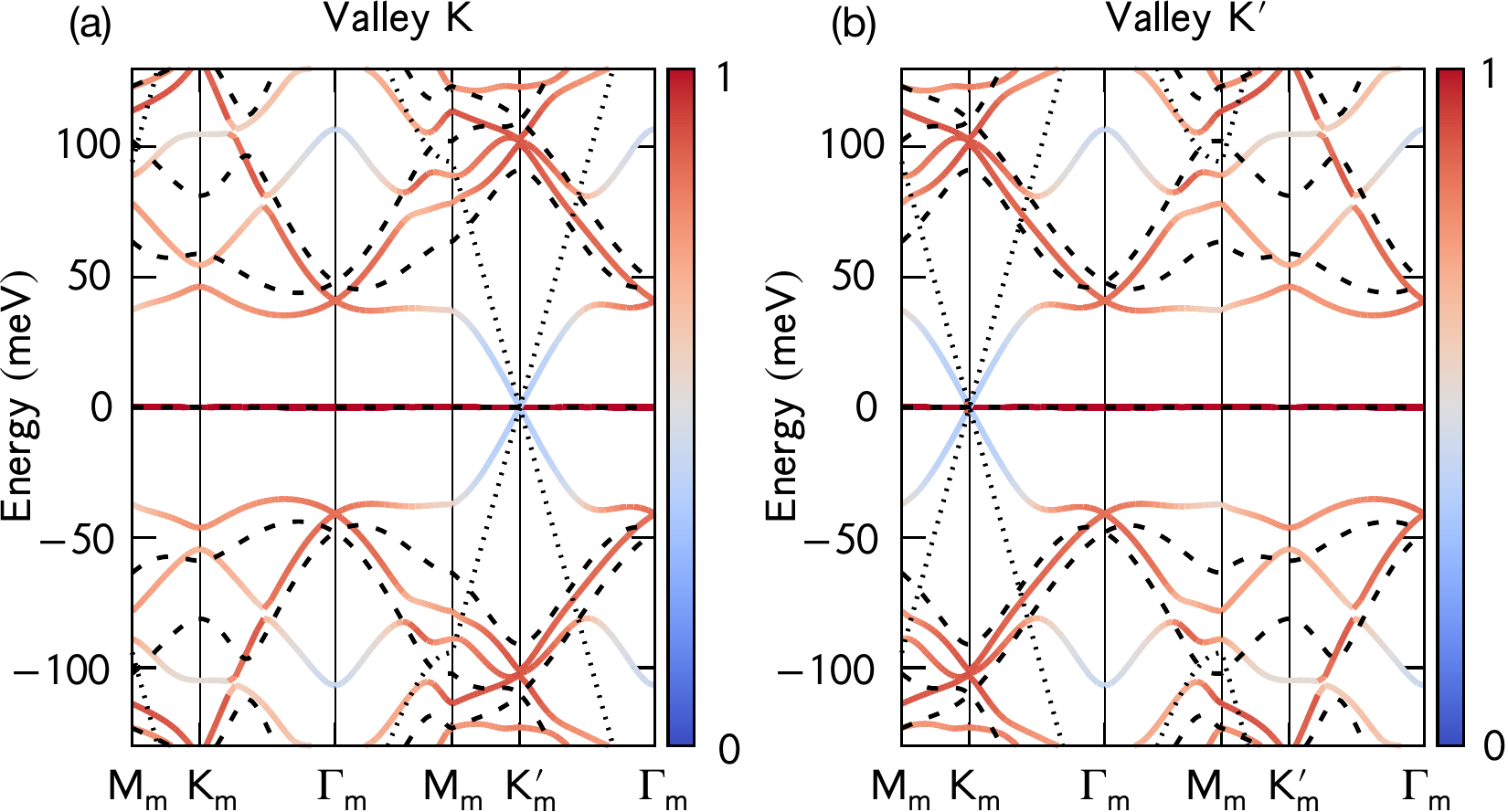}
\caption{Band structure of twisted A-ABA MMG at (a) valley $K$ and (b) valley $K'$  at the magic angle in the chiral limit. The folded band structure of monolayer graphene is shown as dotted lines. The band structure of twisted A-AB graphene is shown as dashed lines. The weight projected onto the twist layers A-AB of A-ABA is indicated by color.}
\label{fig:BS_ABA}
\end{center}
\end{figure}

The band structures of A-ABA at valleys $K$ and $K'$ are shown in Fig.~\ref{fig:BS_ABA}.  The numerical results are consistent with the discussed low-energy effective Hamiltonians Eq.(\ref{eq:HABAprime_eff}). One can observe two flat bands separated from the remote bands as well as a Dirac cone located at $K_m'$ of valley $K$. The Fermi velocity of the Dirac cone in the A-ABA multilayer (blue) is reduced compared to that of monolayer graphene (dotted lines). The results further show that the low-energy spectrum of A-ABA can be regarded as the combination of A-AB (dashed lines) and a monolayer graphene A (dotted lines), 
confirming the validity of our partitioning rules.  
 
\begin{figure}
\begin{center}
\includegraphics[width=8cm]{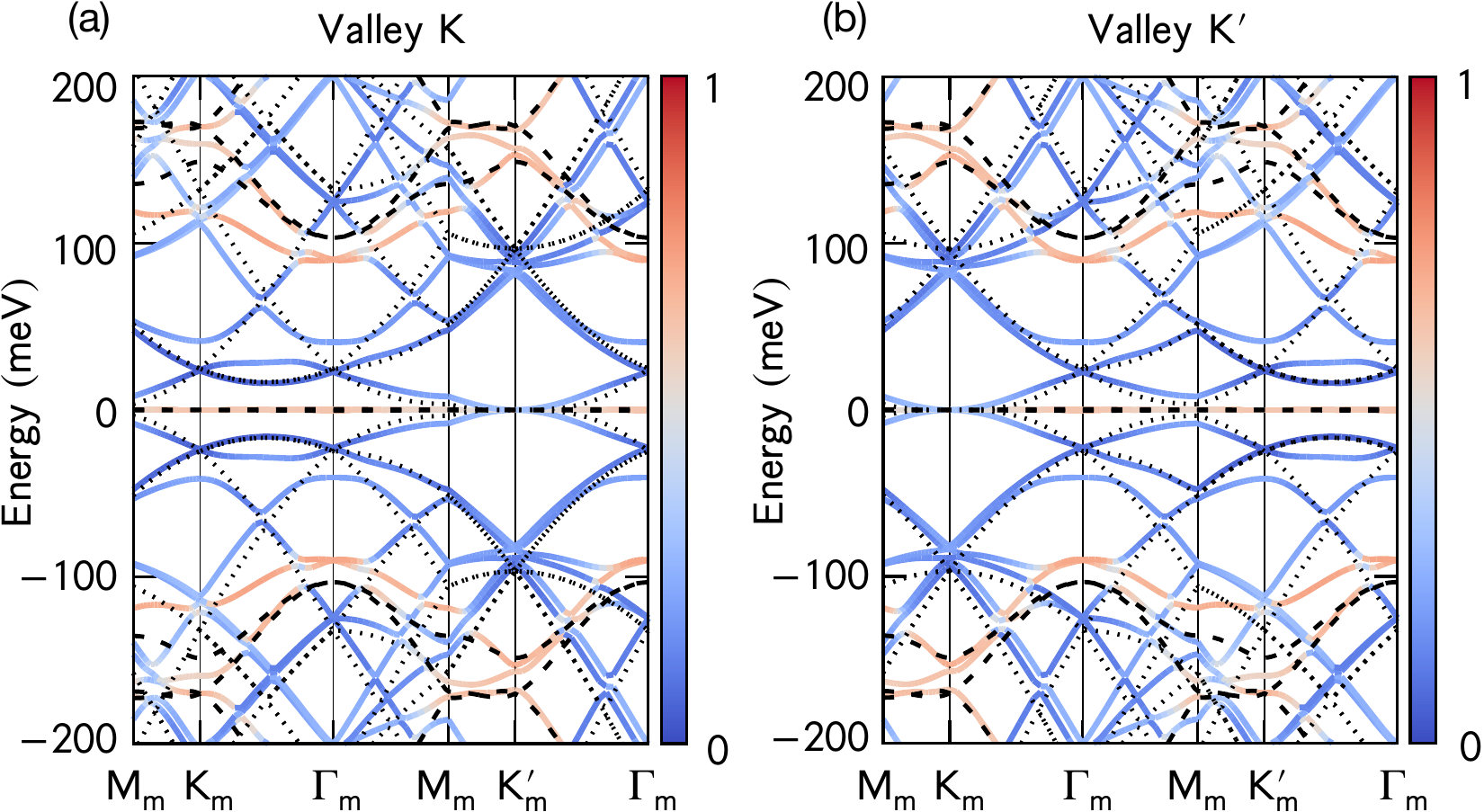}
\caption{Band structure of twisted A-ABAC MMG for (a) valley $K$ and (b)  valley $K'$ at the magic angle in the chiral limit. The folded band structure of BAC trilayer graphene is shown as doted lines. The band structure of TBG is shown as dashed lines. The weight projected onto TBG layers A-A of A-ABAC is indicated by color.}
\label{fig:BS_ABAC}
\end{center}
\end{figure}

Following the same procedure, we project~\cite{supp} the Hamiltonian of the A-ABAC configuration onto the zero-energy states $|LL_{0,\alpha}\rangle$, two isolated site states $|\beta_3\rangle$ and $|\alpha_5\rangle$, and a zero energy state $|\beta_4^-\rangle$ defined in Eq.(\ref{eqn:beta4minus}). The obtained low-energy effective Hamiltonian in the basis of \{\{$|LL_{0,\alpha}\rangle$,  $|\beta_3\rangle$\}, \{$|\alpha_5\rangle$, $|\beta_4^-\rangle$\}\} under the second-order perturbation theory at $K_m'$ ($K_m$) of valley $K$ ($K'$) results in 
\begin{equation}
H_-^{eff}(\bk)=\begin{pmatrix}
0 & 0 \\
0 & 0\end{pmatrix} \oplus
\begin{pmatrix}
0 & - k_-^2/2\tilde{m} \\
- k_+^2/2\tilde{m} & 0
\end{pmatrix} \;
\label{eq:HAABACprime_eff},
\end{equation}
where $\tilde{m}v_F^2=\frac{t\sqrt{t^2+|t_l|^2}}{|t_l|\hbar^2}$. The zero matrix in Eq.(\ref{eq:HAABACprime_eff}) gives two flat bands, and its LDOS is distributed over the $\alpha$ sublattice of TBG layers A-A and the $\beta$ sublattice of the second layer B of the top stacking ABAC. The second matrix depicts a $J=2$ doublet.
The band structures of A-ABAC multilayer at valleys $K$ and $K'$ are shown in Fig.~\ref{fig:BS_ABAC}. These numerical results are also consistent with discussion of low-energy effective Hamiltonians Eq.(\ref{eq:HAABACprime_eff}), showing two flat bands separated from the remote bands and a quadratic dispersion Dirac cone located at $K_m'$ of valley $K$. The results further show that the low-energy spectrum of A-ABAC can be considered as the combination of bands of an A-A bilayer (dashed lines) and a BAC trilayer (dotted lines) graphene, which agrees with our partitioning rules.  

\begin{figure}
\begin{center}
\includegraphics[width=8cm]{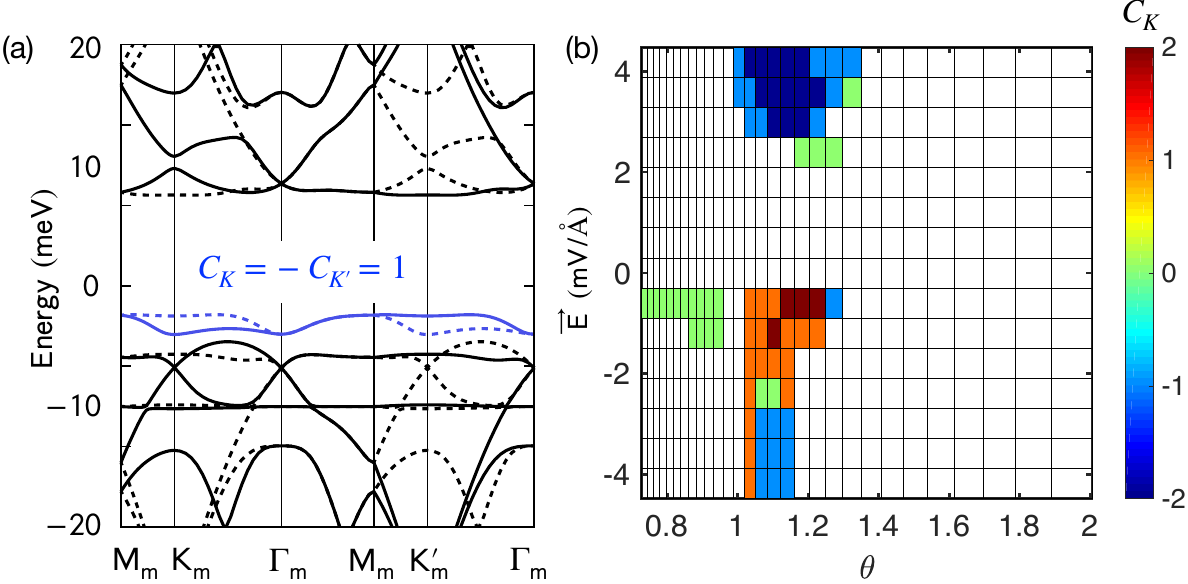}
\caption{(a) Band structure of the A-ABAC twisted multilayer at the magic angle in the chiral limit under vertical electric field $E=4.5$~mV/$\angstrom$. Bands of valleys $K$ and $K'$ are shown as solid and dashed lines, respectively. The Chern number of the isolated flat band in blue in valley $K$ is $C_K=1$. (b) The Chern number of the isolated flat band as a function of twist angle $\theta$  and electric field $E$. 
Blanc cells means that the Chern number is ill-defined.
}
\label{fig:BS_ABAC_Elec}
\end{center}
\end{figure}

Isolated flat bands are often characterized by non-trivial band topology such as the fragile topology~\cite{Song2019,Po2019} or non-zero valley Chern number~\cite{Liu2019a,Liu2019}.
However, in the TMMG systems the $k^J$-dispersive Dirac cones are always entangled with the flat bands, which makes the topological number ill-defined. These bands can be disentangled by applying vertical electrical field, which induces a finite energy gap at the Dirac points \cite{Kumar2011}, thus isolating the flat bands from the other bands. Taking A-ABAC as an example, the band structure in the presence of an external vertically field of 4.5~mV$/\angstrom$ using the chiral limit parameters~\cite{supp} is shown in Fig.~\ref{fig:BS_ABAC_Elec}(a). One can observe a gap opening in the $J=2$ doublet at $K_m'$ of valley $K$, resulting in an isolated flat band with a valley Chern number $C_K=1$. In Fig.~\ref{fig:BS_ABAC_Elec}(b) we show the valley Chern number phase diagram as a function twist angle $\theta$ and electric field $E$ obtained using the realistic  parameters~\cite{supp}. We see that in most of the parameter space the flat bands remain energetically entangled with the quadratic doublet for A-ABAC system with realistic model parameters. However, when the twist angle $\theta\!\sim\!1-1.2^{\circ}$, the flat bands become energetically separated from the quadratic doublet under finite and experimentally accessible electric fields, giving rise to topologically nontrivial flat bands with different valley Chern numbers varying from -2 to 2, which may provide a unique platform to realize interaction driven topological states.

In conclusion, we have formulated partitioning rules that allow understanding the low-energy electronic structure of a broad class of twisted mixed multilayer graphene configurations at the magic angle in the chiral limit. These novel systems contain both the high-mobility Dirac fermions and flat bands that can be disentangled by means of applied electric fields. The flat bands often carry non-zero valley Chern numbers, which has implications for the rational design of twisted graphene multilayers hosting topological electronic phases, e.g. the quantum anomalous Hall insulator. 

\noindent
{\it Note:} During the preparation of our manuscript Ref.~\cite{cao2020} has appeared. This work studies the {\it ab initio} four-band Wannier tight-binding model of TMMG and mentions the chiral decomposition of the low-energy spectrum without detailing the partitioning rules. 

\noindent
{\it Acknowledgments.---}S. Z., Q. W. and O. V Y. acknowledge support by the NCCR Marvel.
First-principles calculations have been performed at the Swiss National Supercomputing Centre (CSCS) under Project No.~s1008 and No.~mr27 and the facilities of Scientific IT and Application Support Center of EPFL. B. X. and J. L. acknowledge the start-up grant of ShanghaiTech University andthe National Key R \& D program of China (grant no. 2020YFA0309601).

\bibliography{refs}
\bibliographystyle{apsrev4-1}

\end{document}